\documentclass[a4paper,11pt]{article}
\usepackage{pos}

\title{Properties of Overlap and Domain Wall Fermions in the 2+1D Thirring Model}

\author*[a]{Jude Worthy}
\author[b]{Simon Hands}

\affiliation[a]{Department of Physics, Swansea University,\\
  Singleton Park, Swansea SA2 8PP, UK}

\affiliation[b]{Department of Mathematical Sciences, University of Liverpool,\\
Liverpool L69 3BX, UK}

\emailAdd{887140@swansea.ac.uk}
\emailAdd{simon.hands@liverpool.ac.uk}

\abstract{We present some results pertaining to partially quenched formulations of the overlap/domain wall operator with the Thirring model in 2+1D. Auxiliary fields are generated with a Shamir domain wall approach and measurements of eigenvalues and condensates are contrasted with different overlap operators. The numerical challenge posed by a non-compact formulation is highlighted, and the effective use of lower accuracy sea fermions is demonstrated.}

\FullConference{%
 The 38th International Symposium on Lattice Field Theory, LATTICE2021
  26th-30th July, 2021
  Zoom/Gather@Massachusetts Institute of Technology
}

\usepackage{braket}

\begin{document}
\maketitle

\section{Introduction}

Quantum field theories are plagued by divergences in their continuum formulations. Regularisation on a lattice renders a theory well-defined. The renormalisation process, when successful, finds fixed-points of such a lattice theory, defined where the correlation length diverges. Each of these is thought to correspond to a well defined theory in the continuum limit. One may alternatively search directly for a critical point through the identification and measurement of an (continuous) order parameter, such as the bilinear condensate, which breaks a symmetry of the theory at the critical point. Further, we may look at the eigenvalues of the Dirac operator. 

This preliminary work continues the exploration of the Thirring model in 2+1D \cite{Hands2015,Hands2016,Hands2019} focusing on (bulk) domain wall fermions \cite{Kap92,Chiu2003,Bor2000}, equivalent to (truncated) overlap fermions \cite{Nar95,Neu98} which permit the continuum $U(2)\to U(1)\times U(1)$ symmetry breaking rather than the $U(1)\times U(1) \to U(1)$ found with staggered fermions. In this framework we look at condensates and eigenvalues, with the eigenvalue analysis of \cite{Nar2021} in mind. Numerical aspects are considered in particular at this stage.

The Euclidean continuum formulation of the Thirring model is given by
\begin{equation}
S[\psi,\bar{\psi}]=\int d^3 x \bar{\psi}(\gamma_\mu \partial_\mu +m)\psi+\frac{g^2}{2}(\bar{\psi}\gamma_\mu\psi)^2
\end{equation}
The self interacting term may be reformulated with an auxiliary field and the usual gauge interacting Dirac term $S[\psi,\bar{\psi}]=S_F[\psi,\bar{\psi},A]+S_G[A]$:
\begin{equation}
S_F[\psi,\bar{\psi},A]=\int d^3 x \bar{\psi}(\gamma_\mu (\partial_\mu +iA_\mu)+m)\psi
\end{equation}
\begin{equation}
S_G[A]=\frac{1}{g^2}\int d^3x A_\mu^2
\end{equation}
This formulation allows Monte Carlo methods to be used in calculations. 

\section{Lattice Dirac Formulations in 2+1D}

Domain wall fermions \cite{Kap92} and subsequently overlap fermions \cite{Nar95} were developed in an attempt to capture the chiral anomaly on the lattice in even dimensions. Domain wall fermions add an extra dimension to the Dirac operator in such a way that chiral fermions are found on the walls, which are separated by the extra dimension. Overlap fermions, formally equivalent to bulk formulations of the domain wall fermions in the infinite limit of the extent of the extra dimension, eliminate the requirement of an extra dimension, and can be expressed compactly utilising the matrix sign function \cite{Neu98}.   

We consider Shamir ($D^S_{DW}$) and Wilson ($D^W_{DW}$) domain wall fermions, both of which are instances of Mobius fermions \cite{Brow2017}. We want to express them in the form $D=D_0+mD_m$. With the extent of the extra dimension set to $L_s=4$ the massless components $D^{S}_{0,DW}$ and $D^{W}_{0,DW}$ may be expressed by
\begin{equation}
D_{0,DW}^{S}=
\begin{pmatrix}
D_W^+ & -P_- & 0 & 0 \\
-P_+ & D_W^+ & -P_- & 0  \\
 0 &  -P_+ & D_W^+ & -P_- \\
0 & 0 & -P_+ & D_W^+ \\
\end{pmatrix},
D_{0,DW}^{W}=
\begin{pmatrix}
D_W^+ & D_W^-P_- & 0 & 0 \\
D_W^-P_+ & D_W^+ & -P_- & 0 \\
 0 &  D_W^-P_+ & D_W^+ & D_W^-P_- \\
0 & 0 & D_W^-P_+ & D_W^+ \\
\end{pmatrix}
\end{equation}
where $D_W^\pm=D_W\pm I$, $D_W$ is the usual Wilson Dirac operator, with a negative mass term, $M$, known as the domain wall height. The usual bare mass term $m$ is incorporated on the domain walls.
\begin{equation}
D_{m1,DW}^{S}=
\begin{pmatrix}
0 & 0 & 0 & P_+ \\
0 & 0 & 0 & 0 \\
0 & 0 & 0 & 0 \\
P_- & 0 & 0 & 0\\
\end{pmatrix},
D_{m1,DW}^{W}=
\begin{pmatrix}
0 & 0 & 0 & -D_W^-P_+ \\
0 & 0 & 0 & 0 \\
0 & 0 & 0 & 0 \\
-D_W^-P_- & 0 & 0 & 0\\
\end{pmatrix}
\end{equation}
In 2+1D we further have the anti-hermitian mass terms
\begin{equation}
D_{m3,DW}^{S}=
\begin{pmatrix}
0 & 0 & 0 & i\gamma_3 P_+ \\
0 & 0 & 0 & 0 \\
0 & 0 & 0 & 0 \\
i\gamma_3 P_- & 0 & 0 & 0\\
\end{pmatrix},
D_{m3,DW}^{W}=
\begin{pmatrix}
0 & 0 & 0 & -iD_W^-P_+\gamma_3 \\
0 & 0 & 0 & 0 \\
0 & 0 & 0 & 0 \\
-i D_W^-P_- \gamma_3 & 0 & 0 & 0\\
\end{pmatrix}
\end{equation}
which eliminate an error term associated with the hermitian mass terms and enable $L_s\to\infty$ measurements to be accurately approximated at significantly lower $L_s$ values \cite{Hands2015}.

For the overlap operator we have
\begin{equation}
\label{EQN::ol}
\begin{split}
D^I_{OL} & = \frac{1+V}{2}+m\frac{1-V}{2}\\
D^{G3}_{OL} & =  \frac{1+V}{2}+im\frac{1-V}{2}\gamma_3\\
\end{split}
\end{equation}
in which $V =\gamma_3 \text{sgn} (H)$ and we consider the kernel $H$ to be either the Shamir kernel $H_S$ or the Wilson kernel $H_W$:
\begin{equation}
\begin{split}
H_W & =\gamma_3 D_W \\
H_S & =\gamma_3 \frac{D_W}{2+D_W} \\
\end{split}
\end{equation}
where $\gamma_3 V \gamma_3 = V^\dagger$ and $D_W\equiv D_W(-M)$ again. In 2+1D the $\gamma_5$ \it may \rm be replaced with $\gamma_3$ as has been done above. 

Even though it is sufficient to show the equivalence of the theories through the equality of the determinants \cite{Ken2006, Hands2016}, since all measurable quantities can be derived from the partition function, and the partition function evaluates to be the determinant, it is nevertheless instructive to see the the precise relation between the full operator matrices. Defining $K_{DW} \equiv C^\dagger D_{DW}^{-1}(1)D_{DW}(m)C$, then
\begin{equation}
K_{DW}=
\begin{pmatrix}
D_{OL}(m) & 0 & 0 & \cdots \\
-(1-m)\triangle_2^R & 1 & 0 &  \\
-(1-m)\triangle_3^R & 0 & \ddots & \\
\end{pmatrix}
=C^\dagger D_{DW}^{-1}(1)D_{DW}(m)C
\end{equation}
For further details the reader is referred to \cite{Brow2017}. Similar relations can be demonstrated for the 2+1D variants above, although it should be noted that introducing Zolatarev coefficients in the Shamir domain wall formulation breaks this relation \cite{Chiu2003}. 

\section{Results}

The auxiliary fields are generated using a rational hybrid monte carlo approach \cite{Cla2004} necessary for exploration with a single dirac field. In all cases these are generated using the Shamir domain wall formulation with coefficients fixed to one, corresponding to the hyperbolic tanh formulation of the overlap operator, and the anti-hermitian mass terms. We then conduct measurements with different overlap operators of different type and $L_s$ value. As such we are generally looking at partially quenched results although we emphasise that when only the $L_s$ value differs in the methodology of the generation of the auxiliary field and the measurements, then the full physics should be achieved in the simulation if both $L_s$ values are large enough. 

\subsection{Kernel Spectra and Condition Number}

\begin{figure}[h]
\begin{center}
\includegraphics[scale=0.5]{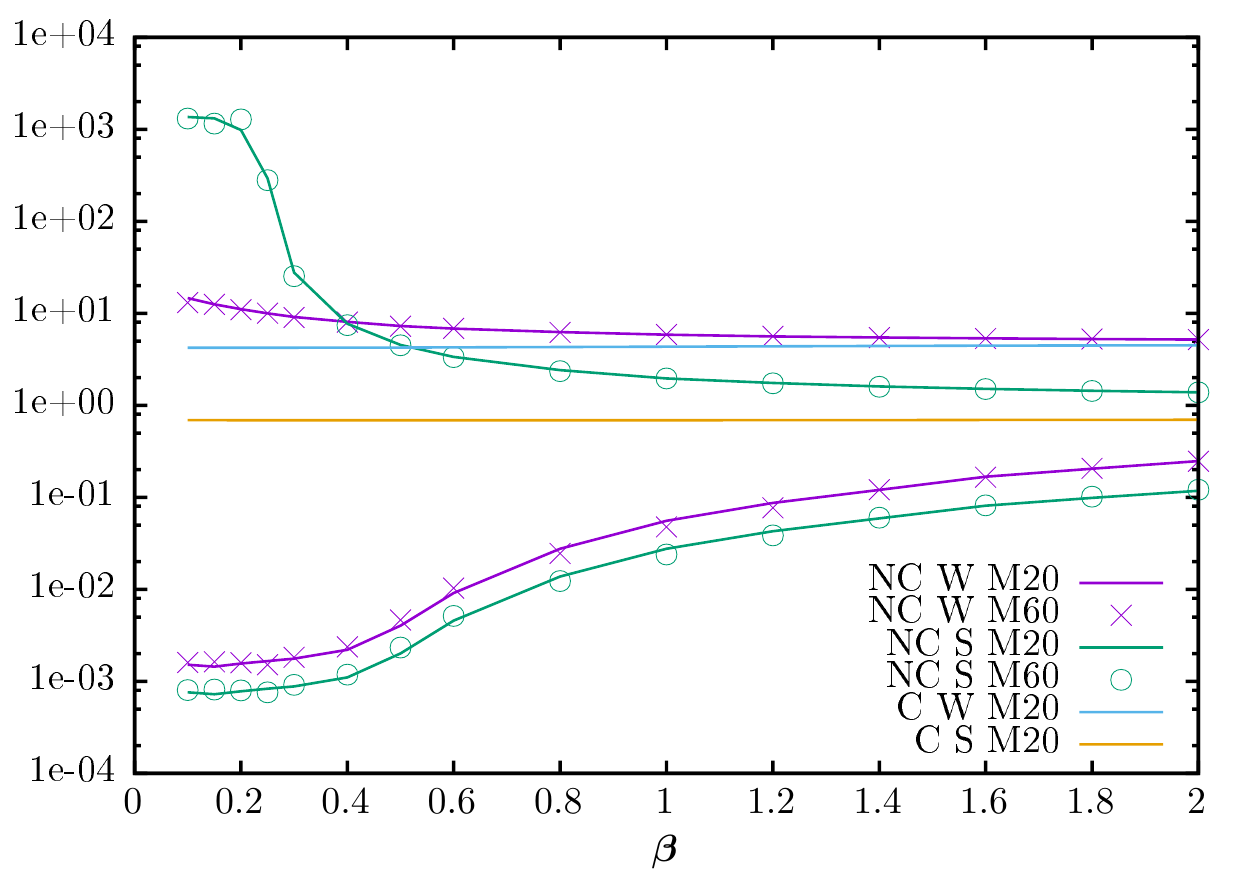}
\includegraphics[scale=0.5]{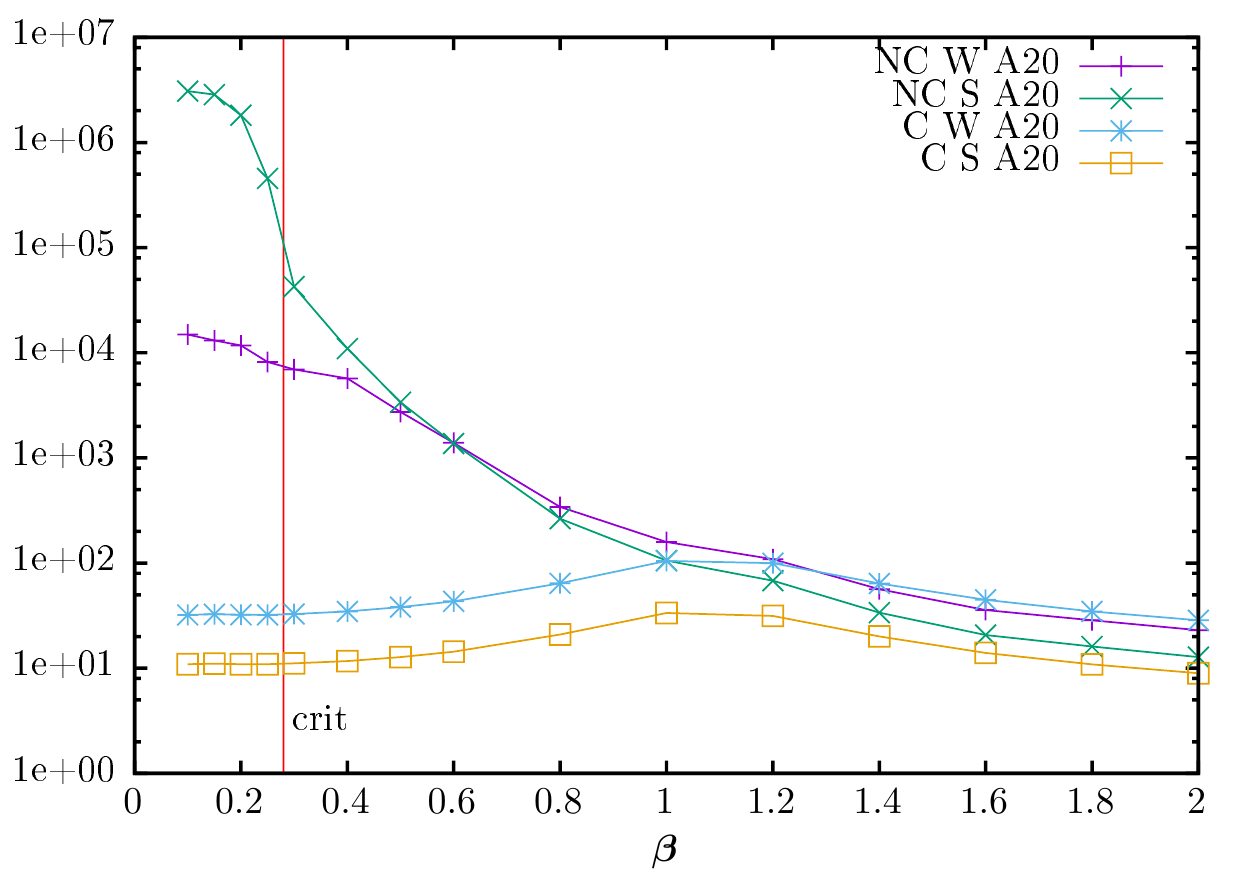}
\caption{LHS: Maximum and minimum eigenvalues for different kernels - non-compact (NC), Wilson(W) and Shamir(S) kernels - produced with $L_s$ values of 20 or 60, plotted against coupling strength $\beta$. A compact (C)  case is also plotted for the maximum eigenvalue only. Note that markers and lines represent different, nearly overlapping, curves in this plot. RHS:The condition number $\kappa(H)$ for compact and non-compact, plotted against coupling strength $\beta=\frac{1}{g^2}$, for different kernels $H$, all using auxiliary (A) fields generated with $L_s=20$.}
\label{FIG::KLsInd}
\end{center}
\end{figure}

Both the Wilson ($H_W=\gamma_3 D_W$) and Shamir ($H_S=\gamma_3 D_W / (2+D_W)$) kernels appear to have minimum and maximum eigenvalues independent of $L_s$, at least above a certain unexplored cutoff. This is shown in the left panel of figure \ref{FIG::KLsInd}. This hints that the auxiliary field structure may be retained even at small values of $L_s$, eliminating the requirement for more arduous calculations in the dynamic creation of the auxiliary fields through the RHMC method. We need to be aware of the spectral range of the kernels, that is the eigenvalue extrema, $\lambda_{\text{min}}(H)$ and $\lambda_{\text{max}}(H)$, and also the condition $\kappa(H)$, to enable a wise choice of Zolotarev or HT parameters. In the latter case a rescaling \cite{Brow2017} may be possible in some cases, but this acceleration technique appears not applicable to the non-compact Shamir formulation due to the unbounded upper eigenvalue. It is similarly unbounded for the non-compact Wilson formulation, but in practice the maximum eigenvalue does not grow so prohibitively, as also indicated in figure \ref{FIG::KLsInd}. A bounded value of a compact case is also shown (artificial as it was created from a non-compact auxiliary field). The condition number is plotted against $\beta=\frac{1}{g^2}$ in the second panel of figure \ref{FIG::KLsInd} and the increased (numerical) challenge of the non-compact formulation around the critical point is in evidence.  

\begin{figure}
\begin{center}
\includegraphics[scale=0.6]{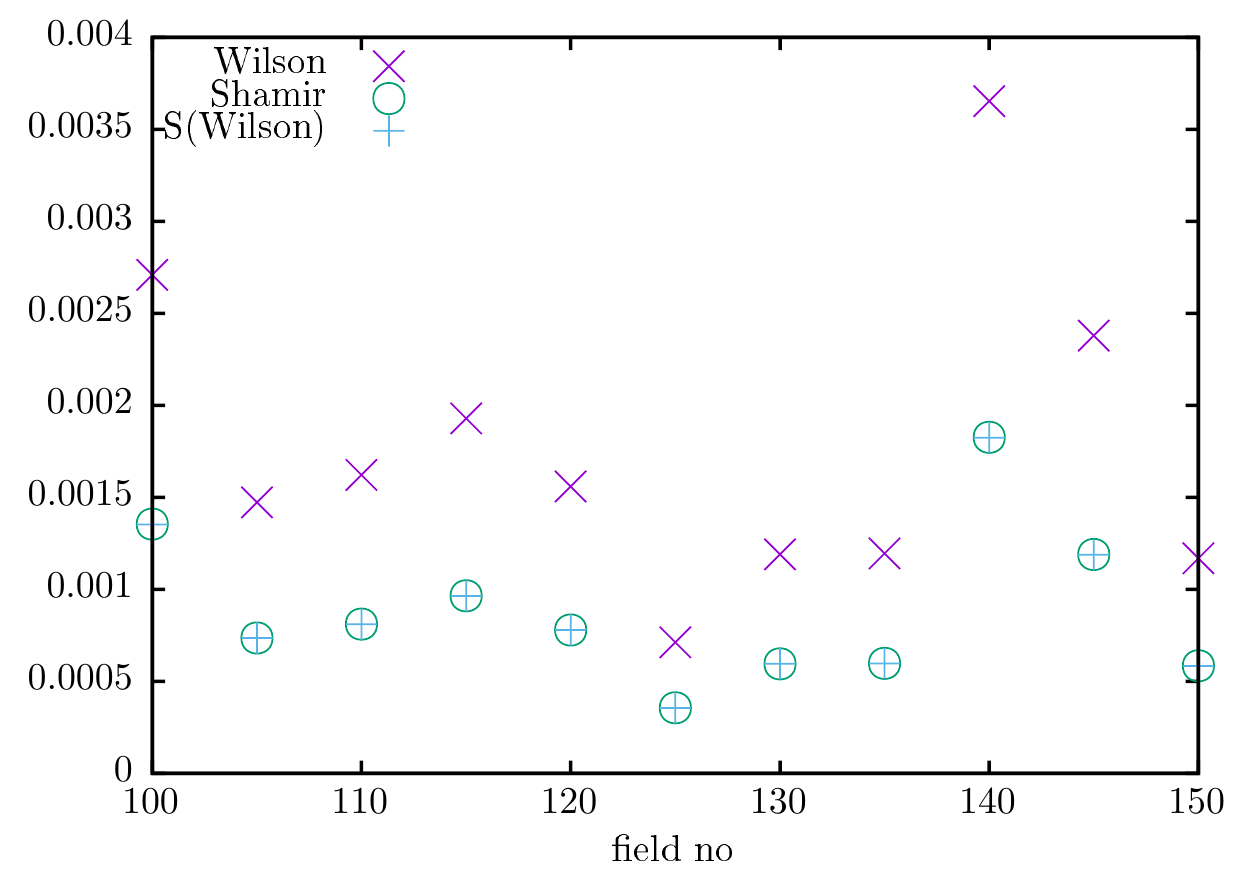}
\caption{The lowest eigenvalues are plotted for the Wilson and Shamir kernels for a range of auxiliary fields. A derived eigenvalue S[Wilson] calculated according to eqn. \ref{EQN::eig} is also plotted, showing the accuracy of the approximation.}
\label{FIG::KSrelW}
\end{center}
\end{figure}

Often, the physics of interest is determined by the smallest eigenvalues. Although it is not formally the case, for the smallest eigenvalues we have the approximation 
\begin{equation}
\label{EQN::eig}
\text{eig}[H_S]\approx\frac{\text{eig}[H_W]}{2+\text{eig}[H_W]}
\end{equation} 
as demonstrated in figure \ref{FIG::KSrelW}. The large eigenvalues have no such approximation. It will be interesting to see if this relation can be exploited in the evaluation of the overlap operator and justifies the interchange of different kernels for sea and valence fermions.

\subsection{Condensate}

The independence of the spectral range on the value of $L_s$ shown in fig \ref{FIG::KLsInd} suggests that it may be sufficient to use auxiliary fields generated with sea fermions using a lower $L_s$ value. This is explored via the bilinear condensate, defined by $C \equiv \frac{\partial \text{ln} Z}{\partial m} = \frac{1}{Z} \braket{\frac{\partial Z_F}{\partial m}}_G$, where $\braket{O}_G \equiv \int \mathcal{D}[U] O[U] \text{exp}(-S_G[U])$
and
\begin{equation}
\frac{\partial Z_F}{\partial m} = \text{Tr} [D^m D^{-1}] \equiv C_F
\end{equation}
For the overlap operators we have
\begin{equation}
\begin{split}
C_{F,OL}^{M1} & =\text{Tr}[\frac{1}{1-m}((D_{OL}^I){-1}-1)] \\
C_{F,OL}^{M3} & = \text{Tr}[\frac{-1}{i\gamma_3+m}((D_{OL}^{G3})^{-1}-1)] \\
\end{split}
\end{equation}
corresponding to the forms given in eqn \ref{EQN::ol}. 
\begin{figure}[h]
\begin{center}
\includegraphics[scale=0.6]{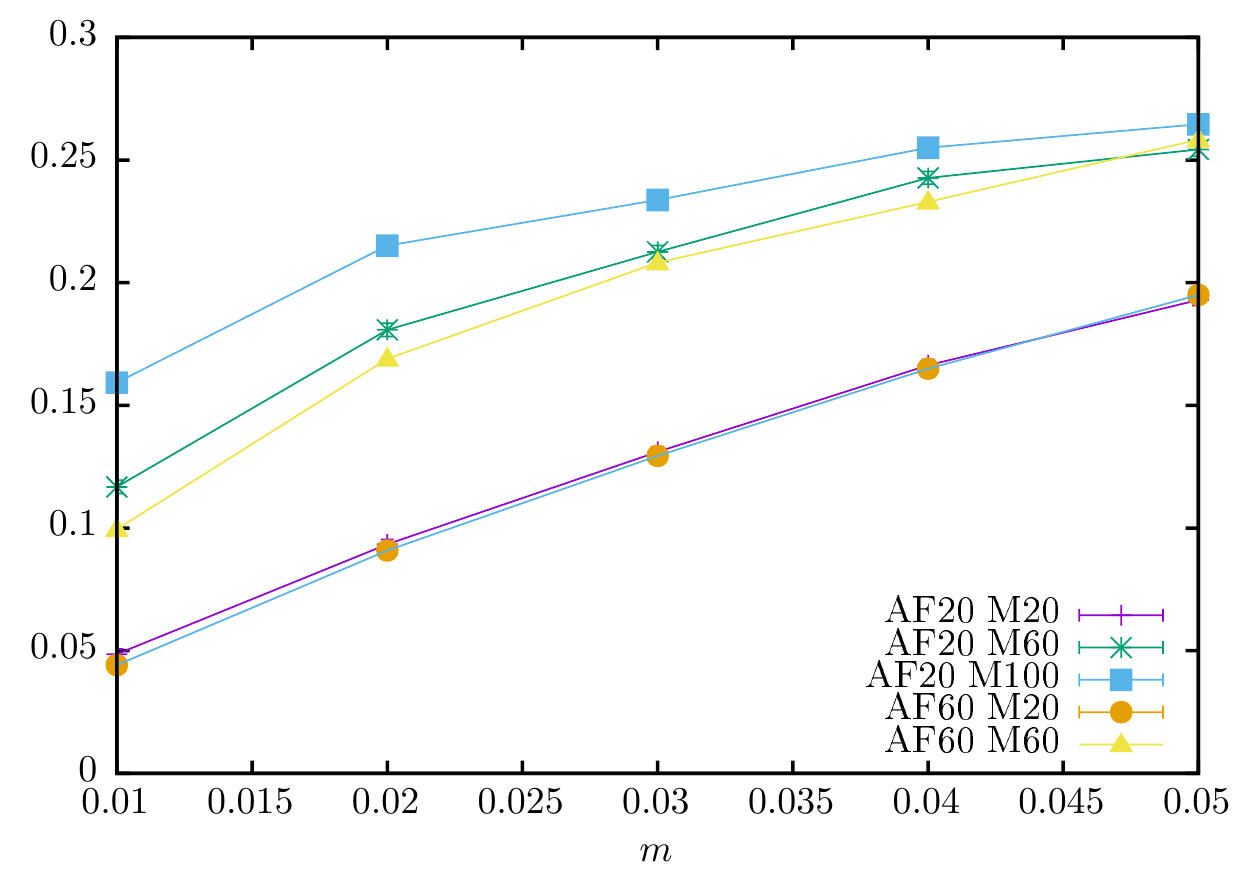}
\caption{Condensate evaluated with $L_s$ values for the generation of the auxiliary fields distinct from the $L_s$ values used for the condensate measurement. The coupling strength is $\beta=0.25$, in the broken phase.}
\label{FIG::cond}
\end{center}
\end{figure}
Figure \ref{FIG::cond} shows condensates over a range of mass values. The curves show different $L_s$ values used for the auxiliary fields and the condensate measurement, both using Shamir kernels, and indicates that the results are predominantly determined by the $L_s$ value of the measurement rather than the auxiliary field, as hoped. There is a clear difference between the results measured with $L_s=60$ with auxiliary fields of $L_s=20$ and $L_s=60$, but this appears to be significantly smaller than the error from not having reached the $L_s$ limit in the measurement. We argue this makes a strong case for the decoupling of $L_s$ in fully dynamic fermions, although what would constitute suitable $L_s$ values would be context dependent. 

\subsection{Overlap Spectra}

We consider the convergence of the lowest eigenvalues of the (Hermitian) overlap operator as found by both forms $D_{OL}^\dagger D_{OL} =2+V+V^\dagger$, and $D_{OL}^\dagger D_{OL} =1+V+V^\dagger+V^\dagger V$ where the former holds for the exact overlap operator, and the latter for the truncated overlap operator, ie $V^\dagger V=1$ as the approximation to the sign function becomes exact. We refer to the second formulation as the alternative formulation, denoted Alt in the plots. The auxiliary fields used are generated with $L_s=20$ Shamir kernels, and zero mass.

\begin{figure}[h]
\begin{center}
\includegraphics[scale=0.5]{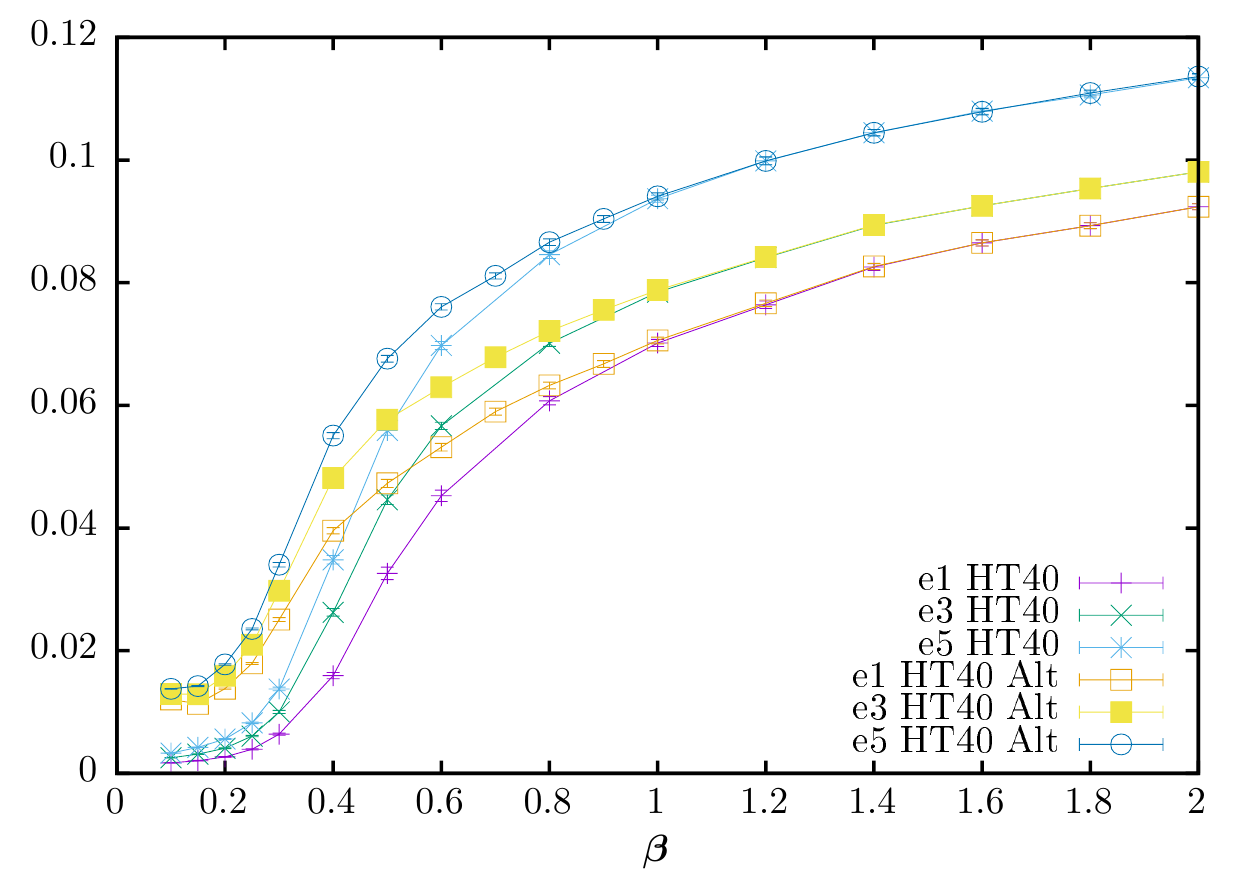}
\includegraphics[scale=0.5]{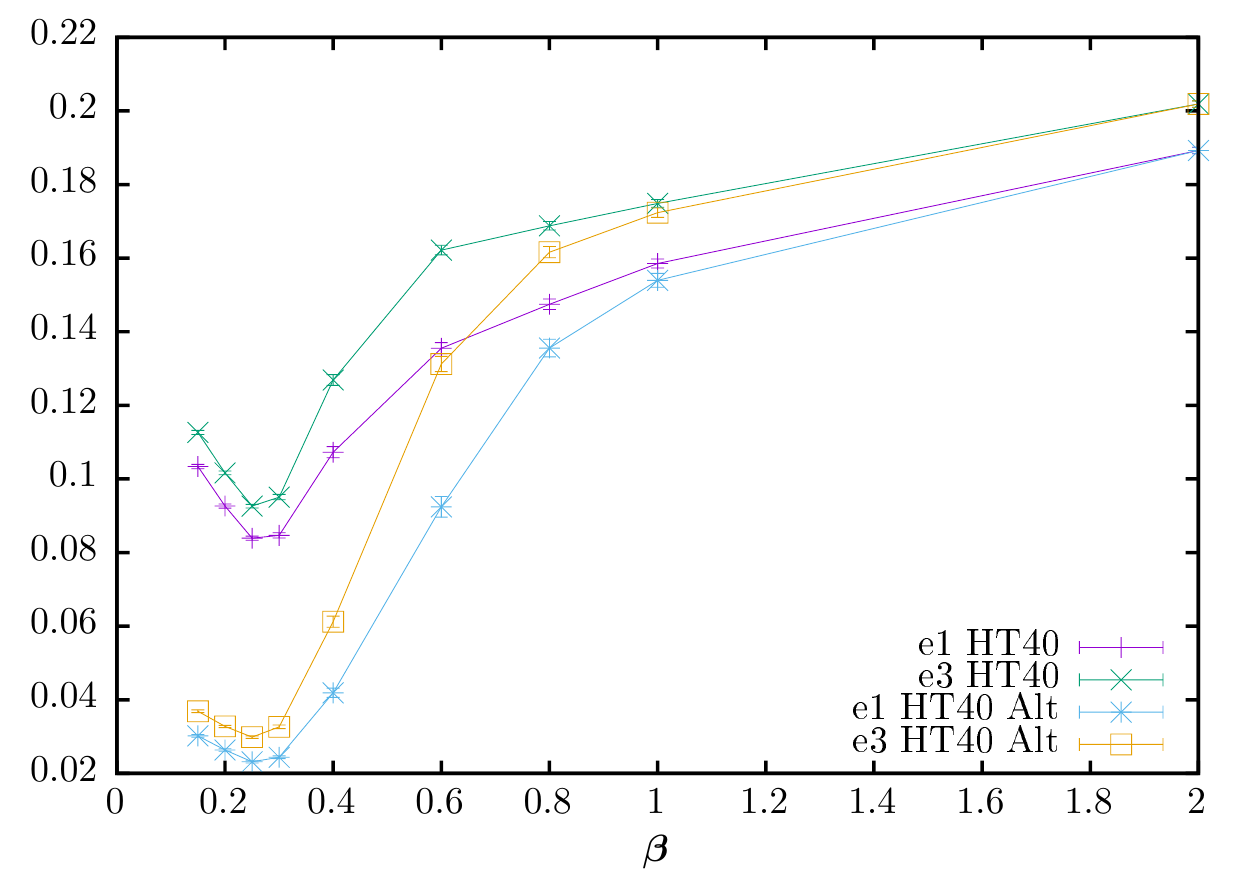}
\caption{LHS: The lowest 3 eigenvalues (e1, e3, e5) of the Wilson overlap operator using the hyperbolic tanh (HT) approximation with $L_s=40$. RHS: The lowest 2 eigenvalues (e1, e3) of the Shamir overlap operator using the hyperbolic tanh (HT) approximation with $L_s=40$.}
\label{FIG::OLWSHT40}
\end{center}
\end{figure}

\begin{figure}[h]
\begin{center}
\includegraphics[scale=0.6]{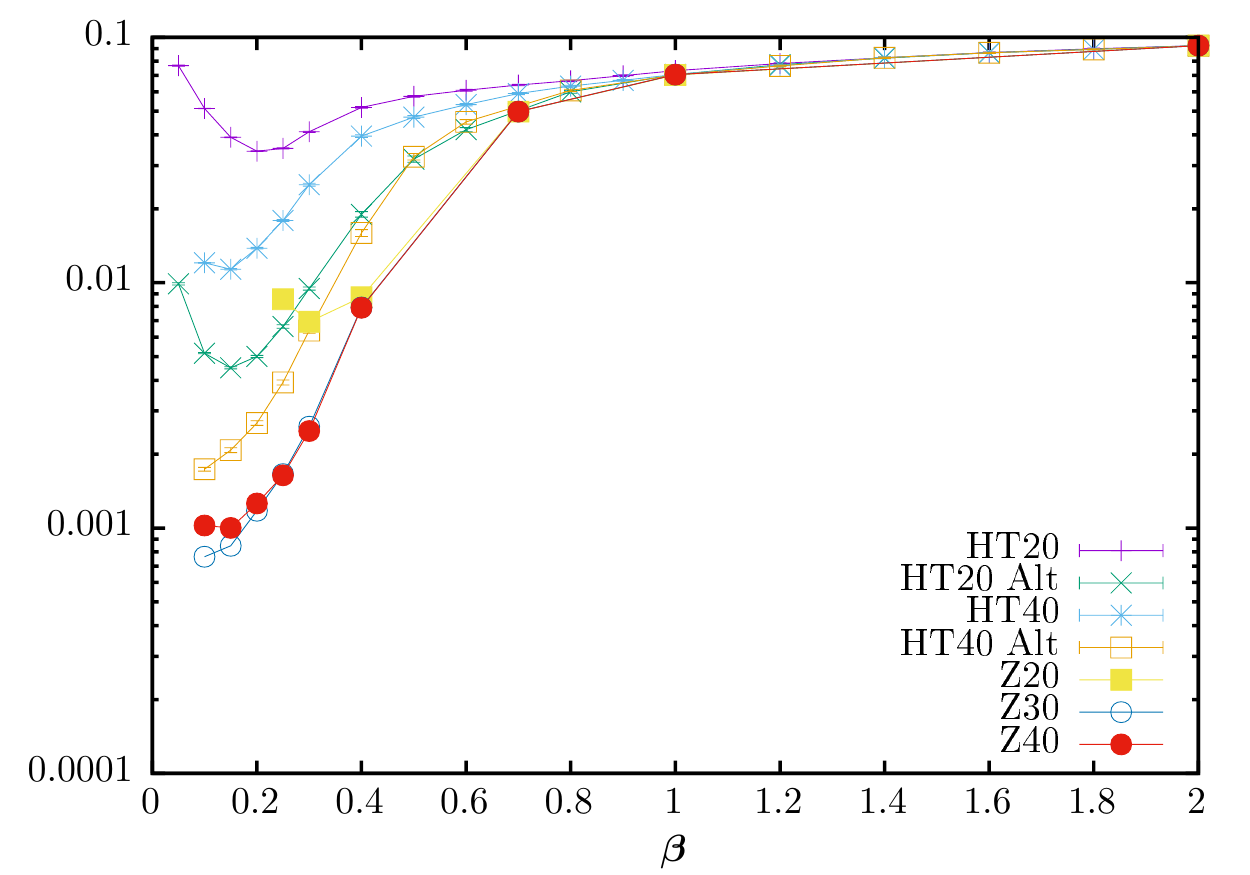}
\caption{The lowest eigenvalue using the Wilson overlap operator using different $L_s$ values and approximation schemes. HT denotes the hyperbolic tanh approximation, and Z denotes the Zolatarev approximation with in the range [1e-4,10].}
\label{FIG::OLW}
\end{center}
\end{figure}

Figure \ref{FIG::OLWSHT40} shows the lowest 1st, 3rd, and 5th lowest eigenvalues since each eigenvalue occurs twice. The first panel shows Wilson kernel results. These very preliminary results may be compared with the quenched compact and non-compact results using the Wilson kernel of \cite{Nar2021}. Their compact case gives an S-curve which we see with our Wilson results, although the cases are not directly comparable since ours is a partially quenched non-compact case. Our Shamir case shows an upturn in the minimum eigenvalue, which perhaps corresponds to the more complex curve found for the non-compact case in \cite{Nar2021}, and is more directly comparable. However, more results need to be obtained. That $L_s$ convergence has not been attained is evidenced by the difference in standard and Alt cases. Figure \ref{FIG::OLW} shows the minimum eigenvalue for Wilson formulations including the Zolotarev formulation. Using an $L_s$ value of 20 with the alternative formulation and the HT approximation gives a better (although far from) converged result at strong coupling than the $L_s=40$ default case. The Zolatarev formulation with $L_s=40$ is visually converged (when comparing with $L_s=50,60$ not shown here), and it remains to be seen if using the alternative formulation will allow for a lower $L_s$ value.

\section{Concluding Remarks}

We have demonstrated the potential for lower $L_s$ valued sea fermions, and highlighted the challenge posed by the unbounded maximum eigenvalue of the non-compact formulation in the overlap kernels in the strongly coupled region. In practical simulations the maximum eigenvalue of the Shamir kernel was significantly higher than for the Wilson kernel, supporting the use of the Wilson kernel in the strongly coupled region if possible. Given the relation between the smallest eigenvalues of the Shamir and Wilson kernels, we hope the partially quenching interchange of kernels is justified, and will continue to investigate in this direction. We will continue examining the eigenvalues of the overlap operators.

\section*{Acknowledgements}

We thank Rajamani Narayanan for helpful discussions and support. This work entailed the use of the Cambridge Service for Data Driven Discovery (CSD3), part
of which is operated by the University of Cambridge Research Computing on behalf of the STFC
DiRAC HPC Facility (www.dirac.ac.uk). The DiRAC component of CSD3 was funded by BEIS
capital funding via STFC capital grants ST/P002307/1 and ST/R002452/1 and STFC operations
grant ST/R00689X/1. DiRAC is part of the National e-Infrastructure. Further work was performed on the Sunbird facility of Supercomputing Wales. The work of JW was supported by an EPSRC studentship, and of SJH by STFC grant ST/L000369/1.

\end{document}